\documentclass[aip,jcp,reprint]{revtex4-1} 
\usepackage{hyperref}
\pdfoutput=1
\usepackage{graphicx}
\usepackage{amssymb}
\usepackage{amsmath}
\usepackage{blkarray}
\usepackage{multirow}
\usepackage{natbib}
\usepackage{epstopdf}
\usepackage{float}

\newcommand{\citen}[1]{%
  \begingroup
    \romannumeral-`\x % remove space at the beginning of \setcitestyle
    \setcitestyle{numbers}%
    \cite{#1}%
  \endgroup   
}

\begin{document}
\title{Overcoming timescale and finite-size limitations to compute nucleation rates from small scale Well Tempered Metadynamics simulations}
%\pt{perhaps replace finite-size by lengthscale to sound resonant with timescale?}
%\ms{well the fact is that length scale is really addressed by the fact that simulations are done with a molecular approach. What we're addressing is really a finite-size issue}
\author{Matteo Salvalaglio}
\email{m.salvalaglio@ucl.ac.uk}
\affiliation{Department of Chemical Engineering, University College London, London WC1E 7JE, United Kingdom}
\author{Pratyush Tiwary}
\affiliation{Department of Chemistry, Columbia University, New York, New York 10027, United States of America}
\author{Giovanni Maria Maggioni}
\affiliation{Institute of Process Engineering, ETH Zurich, CH-8092 Zurich, Switzerland}
\author{Marco Mazzotti}
\affiliation{Institute of Process Engineering, ETH Zurich, CH-8092 Zurich, Switzerland}
\author{Michele Parrinello}
\affiliation{Department of Chemistry and Applied Biosciences, ETH Zurich,CH-8092 Zurich, Switzerland}
\affiliation{Facolt\`a di informatica, Istituto di Scienze Computazionali, Universit\`a della Svizzera Italiana, CH-6900 Lugano, Switzerland}

\begin{abstract}
Condensation of a liquid droplet from a supersaturated vapour phase is initiated by a prototypical nucleation event. As such it is challenging to compute its rate from atomistic molecular dynamics simulations. In fact at realistic supersaturation conditions condensation occurs on time scales that far exceed what can be reached with conventional molecular dynamics methods. Another known problem in this context is the distortion of the free energy profile associated to nucleation due to the small, finite size of typical simulation boxes. In this work the problem of time scale is addressed with a recently developed enhanced sampling method while contextually correcting for finite size effects. We demonstrate our approach by studying the condensation of argon, and showing that characteristic nucleation times of the order of magnitude of hours can be reliably calculated, approaching realistic supersaturation conditions, thus bridging the gap between what standard molecular dynamics simulations can do and real physical systems.
\end{abstract}

\maketitle
\section{Introduction}
Nucleation is the event initiating first order phase transitions in which a small embryo of a thermodynamically stable phase appears within a parent metastable phase. The formation of gas bubbles in a liquid, liquid droplets in a vapour or crystal particles in solution are all examples of nucleation events playing a key roles in a variety of fields ranging from atmospheric physics to pharmaceutical manufacturing.
The small length scale characterising nucleation events renders their direct experimental observation inherently challenging, while providing the ideal playground to apply and develop molecular modelling techniques.   
Despite extensive efforts in the investigation of nucleation phenomena with molecular simulations, the development of a systematic approach to the calculation of nucleation rates from first principles still remains a challenge, due to the very nature of the nucleation phenomena.

In the context of the condensation of a liquid phase from a supersaturated vapour, the nucleation rate $J$ is defined to be the number of liquid droplets formed per unit time and volume.

The formation of a liquid droplet containing $n$ molecules, in a system at constant volume ($V$) and temperature ($T$), is associated to a Helmholtz Free Energy change  $\Delta{F(n)}$.
The maximum of this quantity, $\Delta{F^*}$, corresponding to a critical number of molecules $n_d^*$, constitutes the energy barrier that the system has to overcome in order to undergo the nucleation process. 
The very existence of this barrier determines the key features of nucleation, namely that it is an activated process and a paradigmatic example of a rare event.

The free energy barrier  $\Delta{F^*}$ and the nucleation rate $J$ depend on the thermodynamic driving force, that is typically expressed in terms of the supersaturation S, i.e. the ratio of the actual vapour pressure and the equilibrium vapour pressure. They depend also on temperature, $T$ and on system-specific properties, namely the surface tension between the liquid and vapour phases, the molecular volumes of the two phases, $v_\ell$ and $v_g$, and the specific surface of the newly formed droplets, i.e. $6/d$ for a spherical droplet of diameter $d$.

In a system of volume $V$, at supersaturation $S$, and temperature $T$ the characteristic time of nucleation $\tau$ can be expressed as: 
\begin{equation}
\tau=\frac{1}{J(S,T)\,V}
\label{eq:J}
\end{equation} 

In Eq. \ref{eq:J} it can be readily seen that, at any given condition of temperature and supersaturation, $\tau$, i.e. the average time necessary to observe a nucleation event, is inversely proportional to $V$. This relationship represents a constraint between two key factors determining the computational cost of MD simulations of nucleation events: the number of steps required to observe a nucleation event that needs to be of order of $\tau/\delta{t}$ where $\delta{t}$ is the integration timestep used in MD, and the number of degrees of freedom which is instead proportional to the number of atoms, and thus to the volume $V$ when comparing systems at the same density $\rho=N/V$.
Such a constraint limits the range of conditions that could be directly investigated by small-scale MD simulations to regimes where  $J\geqslant{10^{22}-10^{25}}$ cm$^{-3}$ s$^{-1}$. Up to now, overcoming such limitations has only been possible by using large scale simulations involving millions of atoms and requiring massive computational resources\cite{largeScale}.  

Nucleation is the prototypical example of a rare event, inherently stochastic in nature. When explicitly considering the stochastic character of nucleation $\tau$ represents the expected time for the nucleation of the first liquid droplet from a supersaturated vapour. The \emph{law of rare events} suggests that the first nucleation event can be interpreted as a Poisson process\cite{resnick1992},where the survival probability $P_0$, i.e. the probability that at a given time $t$ there are no droplets in volume V , is:
\begin{equation}
P_0(t)=\exp\left(-\frac{t}{{\tau}}\right)
\label{eq:Survival}
\end{equation} 

Eq. \ref{eq:Survival} highlights how in order to reliably estimate ${\tau}$, the stochastic nature of the nucleation process needs to be explicitly considered, while the reconstruction of the distribution of transition times is pivotal to reliably estimating $J$. In addition to the timescale issues, molecular simulations of nucleation processes suffer from intrinsic finite-size effects that cause a systematic distortion of $\Delta{F(n)}$ and  impact rate calculations\cite{WedekindFiniteSize,Reiss2003,Schmelzer2014,Grossier2009,SalvalaglioPNAS,salvalaglioFaraday2015}.

In this work we propose a systematic approach for the calculation of nucleation rates from small-scale unseeded molecular simulations. The proposed approach is based on recent developments of Well Tempered Metadynamics (WTmetaD)\cite{WellTempered,BarducciReview,ValssonReview2016} that enable for the calculation of transition times distributions from biased simulations. Moreover we develop a systematic correction for the effect of finite-size on rate calculations.

Our method is tested on the paradigmatic case of the nucleation of a liquid argon droplet from a supersaturated argon vapour. The choice of this system has a two-fold aim: the first is to analyse a simple and yet very significant system that allows to draw conclusions general enough to be extended to a wider class of nucleation phenomena. The second is to point out that even in such a simple system, at realistic values of supersaturation the nucleation timescale rapidly grows out of reach of standard MD. 
Moreover choosing argon as a test case allows to benchmark our results against the existing literature based on small-scale unbiased MD simulations\cite{RegueraRate}.

The paper is structured as follows; at first the details of the application of WTmetaD to the calculation of nucleation rates are reported, then an analysis of finite-size effects is carried out. WTmetaD and the finite size correction are then combined to outline a systematic strategy for the calculation of nucleation rates. Finally results are reported and commented.  Unless otherwise noted, the subscript $N$ will be used to refer to relevant quantities in finite-size systems. Such subscript will be dropped whenever referring to their counterpart in macroscopic systems.

\section{Nucleation rates and long timescales}
\label{sec:methods}
\subsection{From metadynamics to dynamics} In this work the acceleration effect associated to WTmetaD has been exploited in order to substantially reduce the simulation time required to observe a nucleation event while simultaneously maintaining the system size small, hence significantly diminishing the overall computational cost.
WTmetaD is conventionally used to compute free energy surfaces in a variety of contexts \cite{BarducciReview,ValssonReview2016}. 
Recently it has been shown that, taking inspiration from conformational flooding \cite{flooding1} and hyperdynamics \cite{voter_prl}, transition times associated to activated events can be efficiently computed from WTmetaD simulations \cite{Tiwary2013}.   
In WTmetaD the simulated system evolves in a transformed time coordinate, $t_{WT}$, due to the application of the history-dependent bias potential $V_B(\xi,t)$ constructed as a function for the collective variable $\xi$\cite{Tiwary2013,BarducciReview}.
As discussed in detail in Ref.~\citen{Tiwary2013}, rate calculations via WTmetaD do not require a converged estimate of free energy profiles, being instead based on the systematic evaluation of the so-called \emph{acceleration factor}, which represents the ratio between the physical time and the meta dynamics time. 

In the context of a nucleation problem, the specific transition time associated to a nucleation event $t_n$ can be computed from the corresponding WTmetaD simulation time $t_{n,WT}$ as: 
\begin{equation}
t_n=t_{n,WT}\langle{\exp\left({\beta{V_{B}(\xi,t)}}\right)}\rangle_{WT}
\label{eq:tau}
\end{equation}
where the term $\langle{\exp\left({\beta{V_{B}(\xi,t)}}\right)}\rangle_{WT}$ is the \emph{acceleration factor} called $\alpha$ in the following. Note that $\beta=1/k_BT$.
Crucial to this procedure is the hypothesis of negligible bias deposition at the transition state. To comply with such hypothesis, the bias potential is constructed through the \emph{infrequent} deposition of potential Gaussians, in a properly defined space of collective variables. The fulfillment of such condition can be checked \emph{a posteriori} using the approach detailed in Ref.~{Salvalaglio2014assessing}. When this is the case, the whole transition time distribution can be recovered from a set of WTmetaD simulations. This approach has been applied to several problems, thus allowing the computation of rates of activated processes such as DNA unfolding \cite{DNAdenat}, and protein-ligand unbinding \cite{UnbindingPNAS,UnbindingTiwary2015,stuckey2016cellular}.

\subsection{A collective variable to describe liquid argon nucleation}

In WTmetaD the bias potential $V_{B}(\xi,t)$ is constructed as a function of a collective variable $\xi$\cite{BarducciReview,DamaConverges,ValssonReview2016}. In this work we choose as collective variable $n$, the total number of liquid argon atoms in the system, corresponding to the typical reaction coordinate used in classical nucleation theory (CNT).
In order to be used in the framework of WTmetaD, $n$ has been expressed as a continuous and differentiable function of the atomic coordinates\cite{LaioPNAS2002,WellTempered,BarducciReview}. 
To this aim the ten Wolde-Frenkel definition \cite{TWFcriterion,RegueraRate} has been applied, in which atoms are considered liquid when they possess a coordination number larger than a threshold value $c_\ell$, that is chosen to be 5. 
The coordination number of each molecule in the system is defined in a continuous and differentiable form through the expression: $c_i=\sum_{j\neq{i}}{f(r_{ij})}$, where $r_{ij}$ is the Cartesian distance between atoms $i$ and $j$ and $f(r_{ij})$ is the switching function: 
\begin{equation}
f(r_{ij})=\frac{1-\left({r_{ij}}/{r_c}\right)^6}{1-\left({r_{ij}}/{r_c}\right)^{12}}
\label{eq:switch1}
\end{equation}
The number of molecules possessing $c_i\geq{c_\ell}$ is thus calculated using the same functional form as in Eq. \ref{eq:switch1}: 
\begin{equation}
n=\sum_{i=1}^N{\frac{1-\left({c_\ell}/{c_i}\right)^6}{1-\left({c_\ell}/{c_i}\right)^{12}}}
\end{equation}
%The bias potential that has been applied during WTmetaD simulations as a function of $n$ is reported in Fig.~S2.  

\subsection{Detecting nucleation events in WTmetaD simulations} 
The biased transition time $t_{n,WT}$ is the simulation time associated with the occurrence of a nucleation event in a WTmetaD simulation. Hence to compute $t_{n,WT}$, it is necessary to reliably detect nucleation events. In our case we apply the approach described in Ref.~\citen{RegueraRate} which is based on the fact that a clear timescale separation exists between the residence time in the supersaturated vapour state and the time necessary for a supercritical nucleus to grow in size. 
The latter phenomenon is orders of magnitude faster than the former, rendering nucleation a rare but fast event. 
As done in Ref.~\citen{RegueraRate}, the nucleation time $t_{n,WT}$ can be directly calculated from the time evolution of $n(t)$, as the simulation time needed to overcome a threshold size of the emerging liquid droplet $\overline{n}$. 
This approach remains valid as long as the threshold size chosen to define the transition criterion is larger than the critical nucleus. In such case the transition time can be safely considered independent of the specific value of $\overline{n}$ \cite{RegueraRate}. %In Fig. \ref{fig:detection} a graphical representation of a typical time series $n(t)$, with  $\overline{n}$, $n^D$ is reported. 

\subsection{Expected nucleation time and nucleation rates} 
\label{sec:nucleation_rates}
As briefly mentioned in the introduction, due to the activated nature of nucleation, its transition time probability distribution is described by the so-called \emph{law of rare events}, and is thus expected to be exponential. The nucleation process, particularly the formation of the first nucleus, i.e. the event that matters at the scale of the MD simulation box, can in fact be modelled by a time-homogeneous Poisson process characterised by a survival probability $P_0(t)=\exp\left({-{t}/{{\tau}_N}}\right)$. Due to its inherent stochasticity, an appropriate sampling of the nucleation times distribution is required to evaluate the expected characteristic nucleation time ${\tau}_N$ \cite{RegueraRate,ArgonRateIsotherms,yuhara2015nucleation}.
In order to compute the expected nucleation time ${\tau}_N$ in a finite-size system, we perform a large number of independent WTmetaD NVT simulations and extract from each of them a nucleation time $t_N$. After this, the survival probability distribution constructed from the $t_N$ values is analysed and its statistical compatibility with a Poisson process quantified. This allows to check whether the conditions under which Eq. \ref{eq:tau} is valid are satisfied \cite{Salvalaglio2014assessing}.
The characteristic nucleation time fitted from the survival probability distribution ${\tau}_N$ is thus used to compute the nucleation rate in the finite size system as  $J_N=\left({\tau}_N{V}\right)^{-1}$, where ${V}$ is the system's volume  \cite{RegueraRate,ArgonRateIsotherms,yuhara2015nucleation}.

\subsection{WTmetaD simulation details}
Transition times were computed from NVT simulations of systems consisting of 512 argon atoms at a constant temperature of 80.7 K. 
A Lennard-Jones potential with $\epsilon=0.99797$ kJ/mol and $\sigma=0.3405$ nm has been adopted to describe the interactions between argon atoms \citep{RegueraRate}. The potential has been truncated, but not shifted, with a cut-off length of $6.75\,\sigma$. The time step for the integration of the equations of motion has been set to $5$ fs \cite{RegueraRate}. These specific conditions were chosen in order to compare our results with those reported in Ref.~\citen{RegueraRate}. The equilibrium vapour pressure ($p_e$) of argon under these conditions is equal to 0.43 bar \cite{baidakov2007,RegueraRate}. 
To carry out rate calculations in a wide supersaturation range we have followed the approach proposed in Ref.~\citen{RegueraRate}, carrying out a series of NVT simulations in cubic boxes of increasing volume, corresponding to supersaturation values ranging from 11.4 to 5.4.
\begin{table}[H]
\caption{WTmetaD simulation setup summary.}
\begin{center}
\begin{tabular}{cccc|ccccc|c}
\hline
 Label &  $S$ & $\l$ & $p$ & $n_{sim}$ & $\sigma$ & $\omega_0$ & $\Delta{t}$ & $\gamma$ & $V_B^0(nl)$\\
         &         &\small{[nm]} & \small{[bar]} &   &    &                    \small{[kJ/mol]} & \small{[ps]}  & &\small{[Y/N]}\\
\hline      
$S_1$ & 11.4   & 10.5 &4.86 &  100 & 0.5/1.0 & 0.01 &  25 & 5 & N\\
$S_2$ & 8.68   & 11.5 & 3.70&  100 & 0.5 & 0.01 & 25 & 5 & N\\                  
$S_3$ & 6.76   & 12.5 & 2.88& 50   & 1.0 & 0.02 & 25 & 5 & N\\
$S_4$ &6.01   & 13.0 & 2.56&  50 & 0.5 & 0.01 & 25 & 5   & Y\\
$S_5$ & 5.36   & 13.5 & 2.28&   50 & 0.5 & 0.01 & 25 &  5  & Y\\
\hline
\label{tab:setup}
\end{tabular} 
\end{center}
\end{table}
In Tab. \ref{tab:setup} setup parameters of the five sets of simulations, namely the supersaturation level $S$, dimension of the simulation box edge $l$, initial pressure $p$, and number of independent simulations per supersaturation level $n_{sim}$ have been reported. In Tab. \ref{tab:setup} we also report the WTmetaD setup parameters, namely width of the deposited Gaussians $\delta$, their initial height $\omega_0$, the deposition stride $\Delta{t}$ and the $\gamma$ factor. For a detailed description of the metadynamics algorithm and parameters the interested reader is invited to check the references\cite{WellTempered,BarducciReview,ValssonReview2016}. In the last column we indicate whether an initial bias potential $V_B^0(n)$ was applied (Y) or not (N) (see the results section for a description of simulations with and without $V_B^0(n)$). 
The highest supersaturation at which we have performed simulations is $S=11.4$, corresponding to the lowest supersaturation at which the standard simulations of Ref.~\cite{RegueraRate} were performed. This allowed us to check that our simulation setup was correctly reproducing nucleation rates both in biased and unbiased simulations.
For each supersaturation the survival probability distribution has been constructed by performing 50 to 100 independent nucleation simulations.  
The bias was updated every 5000 integration steps, with a bias-factor of 5, and a Gaussian height of $1\times10^{-2}$ or $2\times10^{-2}$ kJ/mol. The values of the collective variable $n$ and of the total bias $V_{B}(n,t)$ have been collected every 100 steps. Temperature has been controlled using the Bussi-Donadio-Parrinello thermostat \cite{bussi2007canonical}, with a time constant of $0.1$ ps. WTmetaD simulations were performed with Gromacs 4.6.3 \cite{gromacs} equipped with PLUMED 2.0 \cite{plumed2}.

\section{Nucleation rates and finite size effects}
The nucleation rate of droplets from a vapour can be explicitly derived within classical nucleation theory (CNT) as:
\begin{equation}
J=A\left(\frac{p}{p_e}\right)\exp\left(-\beta\Delta{F^*}\right)
\label{eq:NucleationRate}
\end{equation}
where $A$ is a pre-exponential factor, $p$ is the pressure in the vapour phase, $p_e$ the equilibrium vapour pressure, and  $\Delta{F^*}$ the free energy barrier to nucleation. 
\begin{figure}[H]
\begin{center}
       \includegraphics[width=1.00\columnwidth]{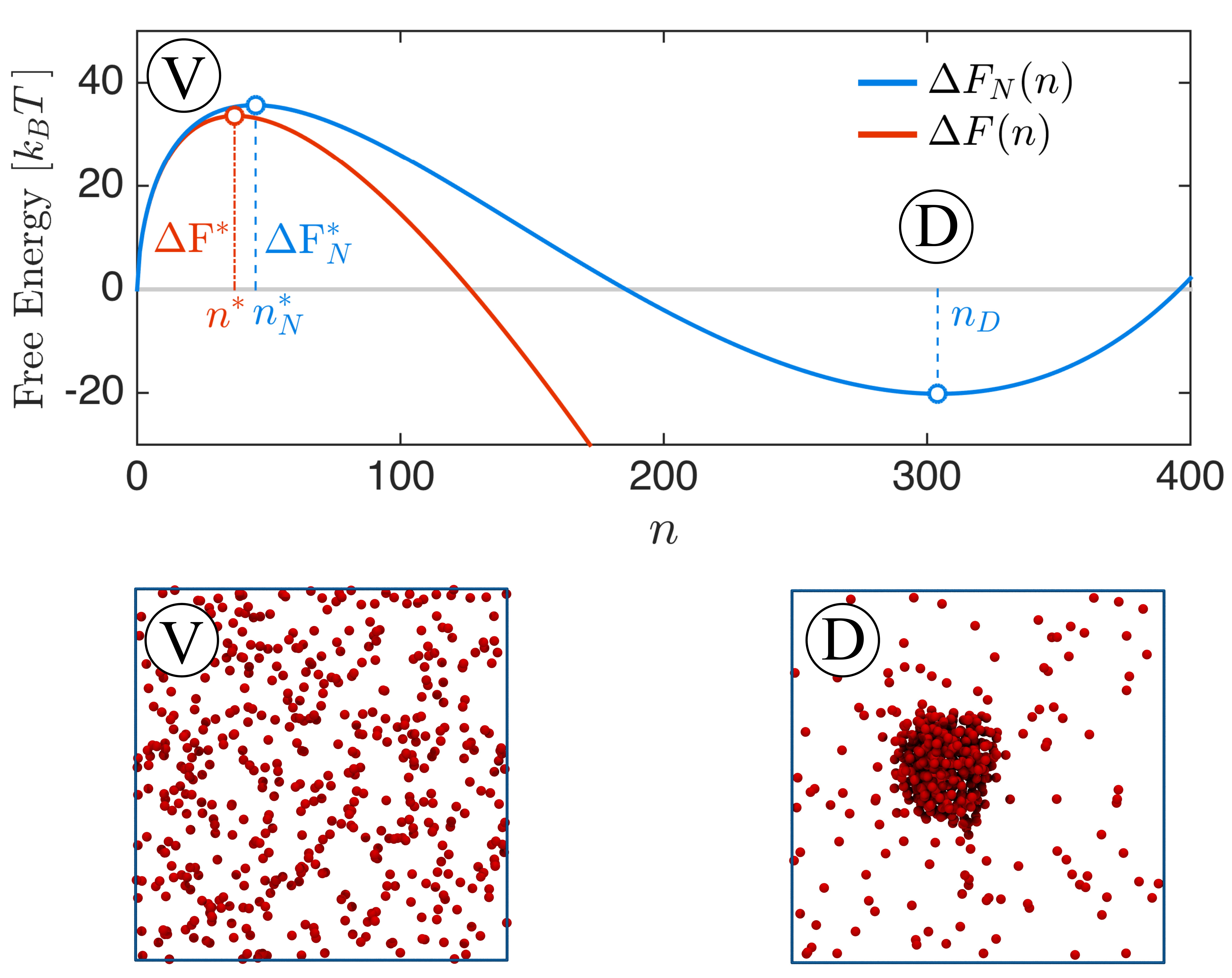}
    \end{center}
   \caption{Nucleation of a liquid argon droplet from supersaturated vapour. \emph{top} Free energy profiles predicted by CNT in an infinitely large system at constant supersaturation ($\Delta{F(n)}$) and for a finite-size, confined system ($\Delta{F_N}(n)$). Both free energy profiles refer to a system of 512 argon atoms, in a volume of 2197 $nm^3$, at supersaturation $S$=6, with a surface energy $\sigma{a}$=9 $k_BT$. \emph{bottom} Representation of the system in its vapour \emph{V} and droplet \emph{D} configurations. }
\label{fig:FES}
\end{figure}

Equation \ref{eq:NucleationRate} can be viewed as the product of two distinct contributions: an energetic part, corresponding to the exponential term, and a 
kinetic one, related to the molecular collisions and given by the pre-exponential term.  
In order to derive a systematic correction to nucleation rates computed from small-scale molecular simulations,  in the following we shall assess the impact of the finite size of the simulation box on both terms. 

\subsection{Free energy of nucleation in a confined system}

\begin{figure*}[ht]
\begin{center}
\includegraphics[width=1.25\columnwidth]{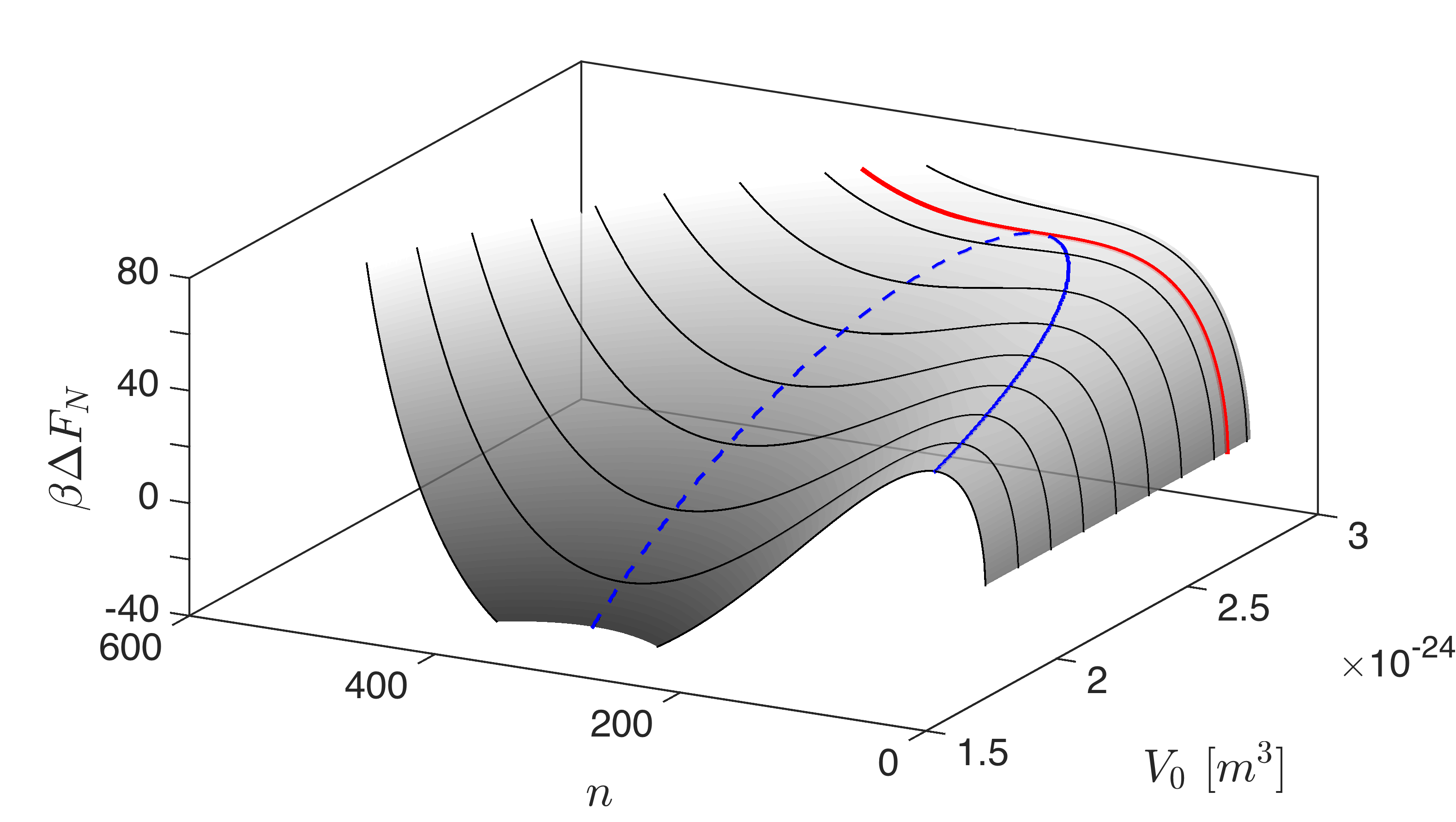}
\end{center}
\caption{Free energy of nucleation in a finite-size system at constant N=512, T=80.7 K, as a function of the system volume V. The solid blue line represents the locus of the maxima of $\Delta{F}_N$, corresponding to the critial nuclei. The dashed blue line represents the locus of the local minima in of $\Delta{F}_N$, representing a stable argon droplet in a finite-size induced equilibrium with the argon vapour. In solid red the $\Delta{F(n)}$ is highlighted, corresponding to the threshold value of V above which $\Delta{F(n)}$ becomes a monotonically increasing function even if $S_0>1$. }
\label{fig:FesN}
\end{figure*}

We define confinement as the impossibility of exchanging atoms between the system and the surrounding environment. Under this definition a $NVT$ simulation box, represents a prototypical confined system, as its total number of atoms $N$ is by definition constant.
As highlighted in several reference works\cite{WedekindFiniteSize,Reiss2003,Schmelzer2014,Grossier2009,SalvalaglioPNAS,salvalaglioFaraday2015}, the free energy change associated with a nucleation process is affected by confinement. In Fig. \ref{fig:FES}, the comparison between nucleation free energies in finite-size (blue) and macroscopic (red) conditions is illustrated together with the representation of typical configurations of Argon atoms in the vapour (V) and liquid droplet (D) states. 
Hereafter we shall summarise how such a distortion in the free energy profile affects nucleation rates\cite{WedekindFiniteSize}. 
We treat the Argon vapour as an ideal gas hence the pressure of the vapour before droplet formation (initial state of the system) is:
\begin{equation}
p_0(N,V,T)=\frac{Nk_BT}{V}
\end{equation}
A liquid spherical embryo of $n$ molecules has a volume $V_d$, and a surface $A_d$, that can be expressed as a function of $n$ as:
\begin{equation}
V_d=nv_\ell
\end{equation}
\begin{equation}
A_d=\left(an\right)^{2/3}
\end{equation}
where $v_\ell$ is the molecular volume in the liquid phase, and $a=6\pi^{1/2}v_\ell$.
After formation of such an embryo (the final state of the system) the vapour pressure attains the following value:
\begin{equation}
p(N,n,V,T)=\frac{\left(N-n\right)k_BT}{V-nv_\ell}
\end{equation}
The Helmholtz free energy change, $\Delta{F}_N$, for the transition from the initial to the final state, i.e. for the formation at constant temperature $T$ and volume $V$ of a $n$-molecule droplet from a vapour consisting initially of $N$ molecules, is given by\cite{Reiss2003}:
\begin{align} \label{eq:DFN}
\Delta{F}_N\left(N,n,V,T\right)=&-n\beta^{-1}\ln{\left(\frac{p}{p_e}\right)} +\gamma\left(an\right)^{2/3} + \nonumber \\ 
&  +N\beta^{-1}\ln{\left(\frac{p_0}{p_e}\right)}+n\left(\beta^{-1}-v_\ell\,p_e\right)
\end{align}
where volume, surface, and pressure effects are accounted for.  A typical $\Delta{F_N(n)}$ profile is reported in Fig. \ref{fig:FES} (a) (blue). Both a local maximum and a local minimum can be identified along $\Delta{F_N(n)}$. The local maximum $\Delta{F}^*_N$ represents the free energy barrier to nucleation in a finite size system.  $\Delta{F}^*_N$ is associated with a critical nucleus size $n^*_N$, that represents the liquid embryo in unstable equilibrium with the surrounding vapour. Such $\Delta{F}^*_N$ value can be computed numerically.
As extensively discussed in Ref.~\citen{WedekindFiniteSize} for an argon vapour and in Refs.~\citen{SalvalaglioPNAS,salvalaglioFaraday2015} for the case of crystal nucleation from solution, there exists a minimum value of the initial supersaturation, $S_0=p_0/p_e>1$, below which the function $\Delta{F_N(n)}$ is monotonically increasing, hence no maximum is present. 
At $N$ and $T$ fixed such a condition defines an upper bound for the volume for which nucleation rates can be computed. In Fig. \ref{fig:FesN} $\Delta{F_N(n)}$ is plotted as a function of the system volume, highlighting the critical conditions associated with the transition to a monotonically increasing function. 
Contrary to the analysis so far, Classical Nucleation Theory (CNT) deals with infinitely large systems, where the formation of the liquid droplet has a negligible effect on the state
of the surrounding phase. The corresponding Helmholtz free energy change, $\Delta{F(n)}$, for the formation of a $n$-molecule embryo at temperature $T$ and supersaturation $S_0$ can be obtained by taking the limit of Eq. \ref{eq:DFN} with $V$ and $N$ approaching infinity, and their ratio remaining constant. 
Under these conditions $p=p_0$ and:
\begin{equation} 
 \Delta{F(n)}=-nk_BT\ln{S_0} +\gamma\left(an\right)^{2/3} 
\label{eq:CNT}
\end{equation}

For an infinitely large system, in which Eq. \ref{eq:CNT} holds the free energy barrier $\Delta{F}^*$ can be computed analytically as \cite{Kashchiev}:
\begin{equation}
\Delta{F^*}=-\frac{4\beta^3\gamma^3a^2}{27\left(\ln{S_0}\right)^2}
\end{equation} 

It is worth noticing that $\Delta{F^*_N}$, which is the nucleation free energy barrier in a confined system at the same conditions of $T$ and $S_0$, is strictly larger than $\Delta{F^*}$.

\subsection{Macroscopic nucleation rates from finite size calculations}

\begin{figure*}
\begin{center}
\includegraphics[width=2.0\columnwidth]{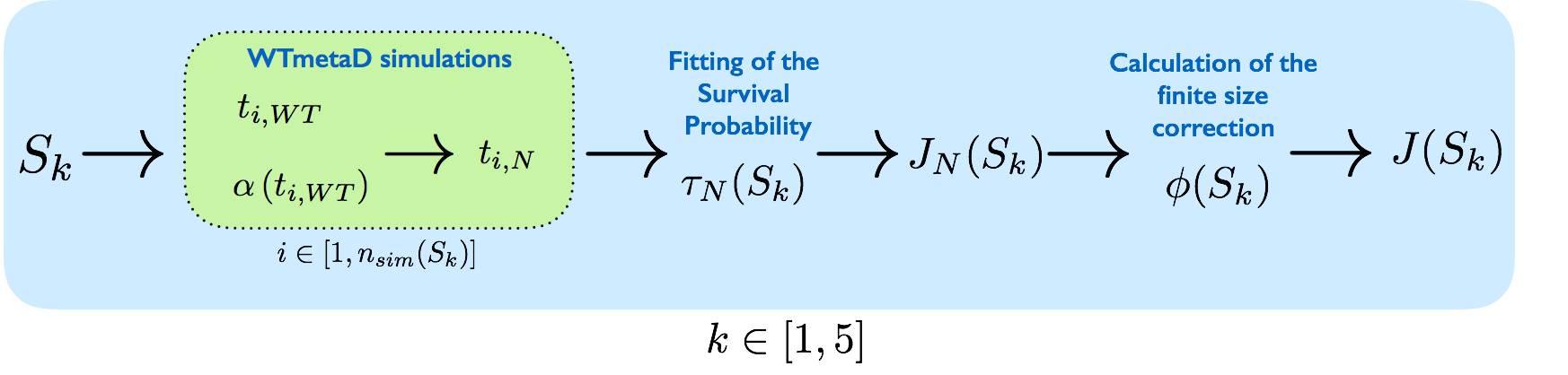}
\end{center}
\caption{Workflow summary for the calcuations of macroscopic nucleation rates from finite size NVT WTmetaD nucleation simulations.}
\label{fig:Workflow}
\end{figure*}

In order to compute a correction term associated to the confinement effect we follow the approach of Ref.~\citen{WedekindFiniteSize}, and define a factor $\phi$ as the ratio between the nucleation rate in macroscopic conditions $J$ and in a finite sized confined system $J_N$.
In analogy with Eq. \ref{eq:NucleationRate}, the nucleation rate in a confined system $J_N$ can be written as\cite{Reiss2003} 
\begin{equation}
J_N=A_NS\exp\left(-\beta\Delta{F^*_N}\right)
\label{eq:JN}
\end{equation}
Since the system supersaturation in a finite-size simulation before a supercritical nucleus forms is essentially the same as in a macroscopic system, $S=S_0$, $\phi$ reduces to\cite{WedekindFiniteSize}: 
\begin{equation}
\phi=\frac{J}{J_N}=\frac{A}{A_N}\exp\left(\beta(\Delta{F}^*_N-\Delta{F}^*)\right)
\label{eq:ratio}
\end{equation}

Eq. \ref{eq:ratio} provides a working principle to obtain nucleation rates in macroscopic systems from finite size NVT simulations as: 
\begin{equation}
J=\phi{J_N}
\label{eq:ratio2}
\end{equation}

As reported in Ref.~\citen{WedekindFiniteSize}, due to the exponential dependence on the strictly positive quantity $\Delta{F}^*_N-\Delta{F}^*$, the dominating term in Eq. \ref{eq:ratio} is $\exp\left(\beta(\Delta{F}^*_N-\Delta{F}^*)\right)$, whereas the pre-exponential term $A/A_N$ in Eq. \ref{eq:ratio} is expected to play a secondary role. In order to identify key contributions to  $A/A_N$, its dependence on finite size is discussed in the following. 

Within the framework of CNT, $A$ is typically expressed as\cite{RegueraRate,Kashchiev}
\begin{equation}
A=Zf^*\frac{p_e}{k_BT}
\end{equation}

where $Z$ is the Zeldovich factor given by\cite{zeldovich1943}:
\begin{equation}
Z=\sqrt{\frac{\vert{\frac{d^2\Delta{F}(n)}{dn^2}}\vert_{n=n^*}}{2\pi{k_BT}}}
\label{eq:Zeldovich}
\end{equation} 
and $f^*$ the rate of attachment of molecules to the critical cluster. 
Since nucleation of a droplet from its vapour is a process controlled by direct impingement \cite{Kashchiev}, the attachment rate $f^*$ is derived from the kinetic theory of gases as \cite{Kashchiev,wedekind2007nucleation}:  
\begin{equation}
f^*=c(n^*)\frac{p}{\sqrt{2\pi{m}k_BT}}
\end{equation}
where $c(n^*)=\sqrt[3]{(36\pi{v_\ell}^2)}(n^*)^{2/3}$ is the surface area of the critical cluster, $v_\ell$ is the volume per molecule in the liquid phase and $p$ the pressure.

The attachment frequencies $f^*$ and $f^*_N$ differ due to two reasons. The first is that the critical nucleus size in finite size simulations $n^*_N$ is strictly larger than the critical nucleus in the corresponding infinite case $n^*$\cite{Reiss2003}. The second reason is that the vapour pressure acting on the critical nucleus in finite size systems $p_N^*=(N-n_N^*)/((V-n_N^*v_\ell)k_BT)$ is always smaller than its corresponding value for a system at macroscopic conditions $p=N/(Vk_BT)$. 

The Zeldovich factors $Z$ and $Z_N$ are instead expected to differ due to the fact that the curvature of the free energy profile in the region around its maximum is affected by finite size, see Fig. 1 for example. 
The extent of the contribution of the $\exp\left(\beta(\Delta{F}^*_N-\Delta{F}^*)\right)$, $f^*/f^*_N$, and $Z/Z_N$ terms to $\phi$ is discussed in the Results section.

The correction factor $\phi$ depends on quantities that can be directly calculated from $\Delta{F(n)}_N$ and $\Delta{F(n)}$ such as: $\Delta{F}_N^*$, $\Delta{F}^*$, $Z_N$, $Z$, $n^*_N$, and $n^*$.
Both $\Delta{F(n)}$ and $\Delta{F(n)}_N$ can be respectively computed from Eq. \ref{eq:CNT} and \ref{eq:DFN}, once the surface tension $\gamma$ is known. 

The surface tension $\gamma$ is obtained by fitting Eq. \ref{eq:JN},  in which the pre-exponential term $A_N$ in considered supersaturation-independent, on the $J_N$ values obtained as a function of supersaturation.

\section{Workflow Summary}

In Fig. \ref{fig:Workflow} the workflow for the calculation of nucleation rates from small-scale finite size NVT simulations has been summarised. 
The calculation procedure can be outlined as follows: 
\begin{enumerate}
\item A set of WTmetaD simulations is carried out for multiple supersaturation levels. Supersaturation is imposed by defining the system volume while keeping constant the number of molecules $N$ and the temperature $T$. 
\item Applying the criterion for the identification of nucleation events proposed in Ref.~\citen{RegueraRate}, the WTmetaD transition time $t_{WT}$ and the corresponding acceleration factor $\alpha$ are calculated from each WTmetaD simulation. 
\item The physical transition time associated to each nucleation event is computed using Eq. \ref{eq:tau}.
\item The transition times obtained for each supersaturation value are used to fit the survival probability distribution, and compute the average nucleation time $\tau_N$ for each finite size system at volume $V$ and supersaturation $S$.
\item Average nucleation times $\tau_N$ are converted to finite-size nucleation rates using $J_N=1/\left(\tau_NV\right)$\cite{RegueraRate}. 
\item The finite-size nucleation rates are used to fit Eq. \ref{eq:JN}. The fitting parameter is the surface tension $\gamma$, which is used to compute $\Delta{F}*$, $\Delta{F}_N^*$, $n^*$, $n^*_N$, $Z$, $Z_N$, and thus the correction factor $\phi$. 
\item Eq. \ref{eq:ratio2} is used to compute the nucleation rate in macroscopic conditions $J$.
\end{enumerate}

\section{Results}

\subsection{WTmetaD simulations}
\begin{figure}[H]
     \begin{center}
       \includegraphics[width=1.0\columnwidth]{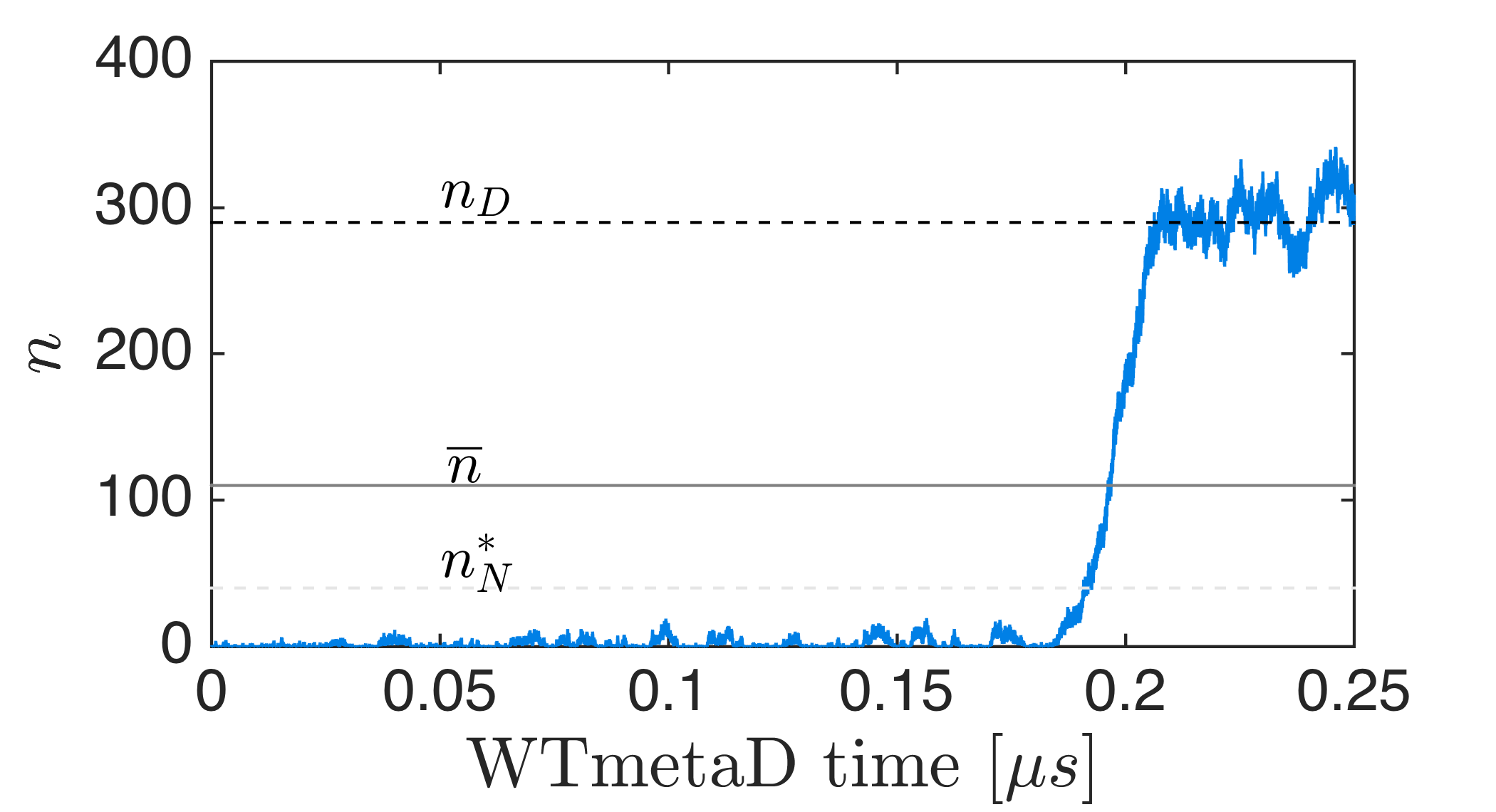}
    \end{center}
    \caption{Time series of the collective variable $n(t)$ obtained from a typical simulation. The value of typical $\overline{n}$, $n_N^*$, and $n_D$ have been highlighted on the plot.}
\label{fig:detection}
\end{figure}

The time evolution of the number of liquid-like argon atoms $n$ in a typical WTmetaD simulation is reported in Fig. \ref{fig:detection}, where the nucleation event can be clearly identified as the rapid transition from $n$ values fluctuating close to zero, to $n$ values fluctuating around a positive value $n_D$. The final state corresponds to a finite-sized droplet stabilised by finite size effects corresponding to the local minimum in free energy shown in Fig. \ref{fig:FES} (blue curve). 

\begin{figure}[H]
\includegraphics[width=1.0\columnwidth]{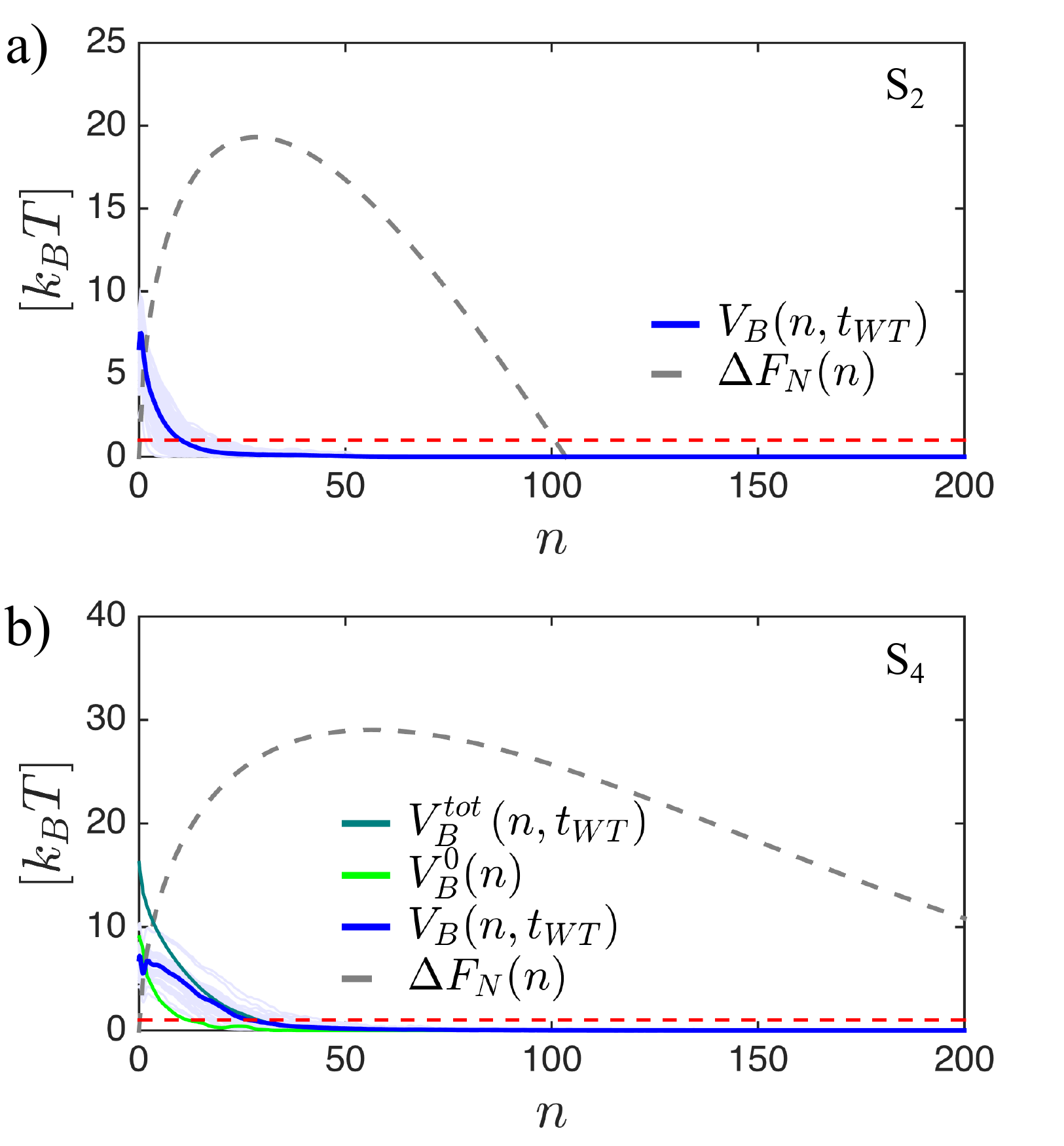}
\caption{a) Simulation set $S_2$: WTmetaD bias potential at transition time, $V_B(n,t_{WT})$. b) Simulation set $S_4$: initial static bias potential $V_0(n)$, WTmetaD bias potential at transition time $V_B(n,t_{WT})$, and total bias at transition time $V_B^{tot}(n_,t_{WT})=V_B^0(n)+V_B(n,t_{WT})$. For comparison nucleation free energy profiles $\Delta{F_N(n)}$ have been reported and the $k_BT$ level has been highlighted in red.}
\label{fig:Bias}
\end{figure}

It can be seen that the lifetime of the supersaturated vapour state in the WTmetaD simulation is much larger than the transition time associated with the nucleation event driving the system into the stable state characterised by $n=n_D$. 
Such a difference becomes exponentially large when the WTmetaD time is rescaled to real time according to Eq. \ref{eq:tau}\cite{Tiwary2013}, neatly highlighting the timescale separation characteristic of the nucleation problem. 
During WTmetaD simulations a repulsive bias potential is adaptively constructed with an infrequent deposition of Gaussians\cite{WellTempered,Tiwary2013}. In order to speed up the adaptive construction of the bias for the two slowest cases ($S_4$ and $S_5$ in Tab . \ref{tab:setup}), in addition to the WTmetaD bias $V_B(n,t)$, we apply a static bias $V_B^0(n)$ constructed from a preliminary WTmetaD simulation. In Fig. \ref{fig:Bias} the total bias potential is reported for supersaturation levels $S_2$ and $S_5$, which are characterised by the  absence and presence of an initial bias $V_B^0(n)$, respectively. In all cases in the region of the maximum of $\Delta_N{F(n)}$, the total bias applied $V_B^{tot}(n)$ decays to values smaller than $k_BT$, in agreement with the hypotheses of negligible bias deposition at the transition state invoked to carry out rate calculations from WTmetaD\cite{Tiwary2013,Salvalaglio2014assessing}.
\begin{figure*}[t]
\includegraphics[width=2.0\columnwidth]{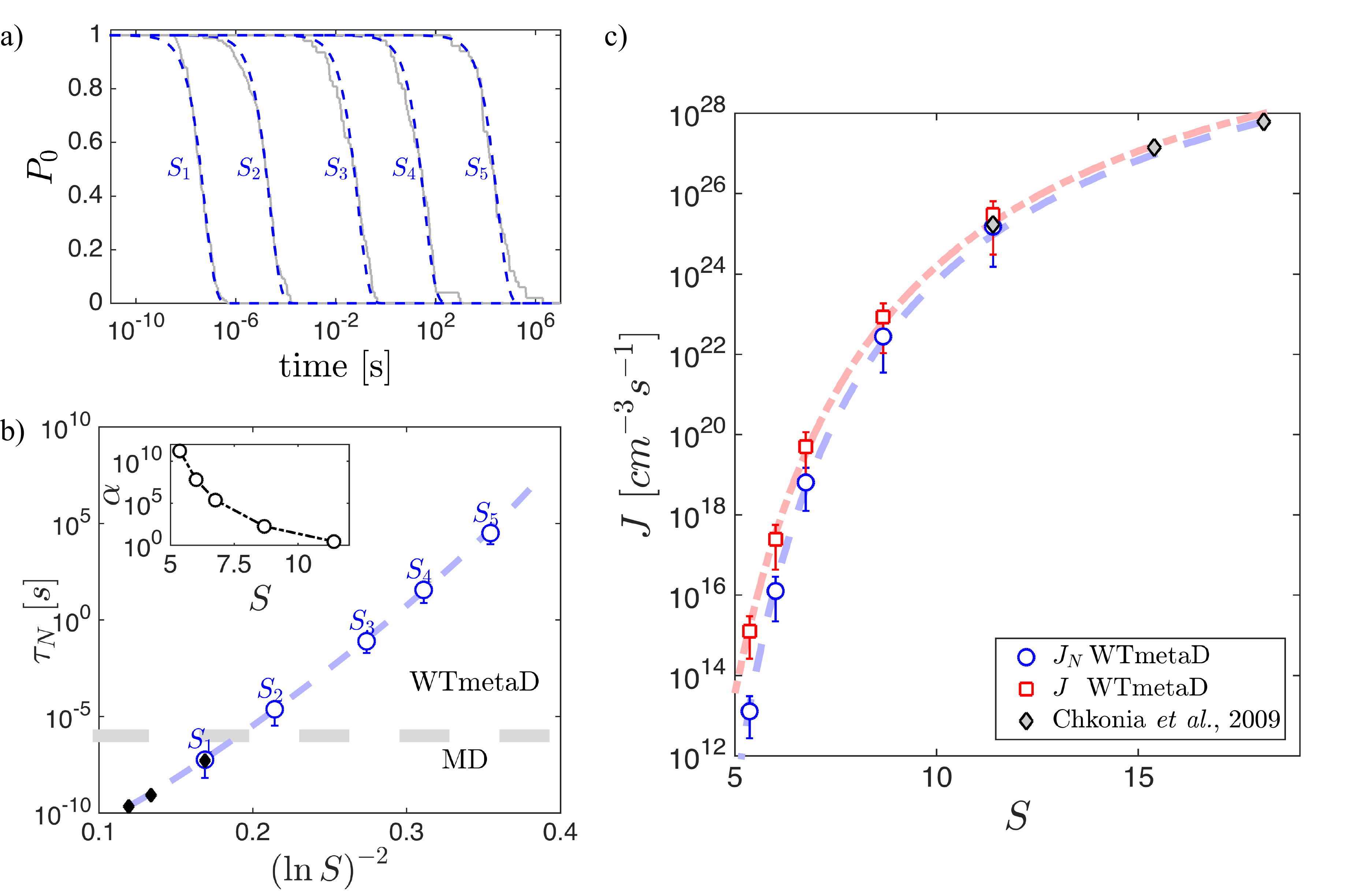}
   \caption{\emph{a)} Survival probability distributions obtained for different supersaturations ($S_1$=11.4, $S_2$=8.7, $S_3$=6.8, $S_4$=6.0, $S_5$=5.4). \emph{b)} Average nucleation times in finite size  systems $\tau_N$. The average acceleration factor $\alpha$ as a function of supersaturation is displayed as an inset. \emph{c)} Nucleation rates calculated from finite-sized WTmetaD simulations ($J_N$) and rescaled to the macroscopic limit $J$. The blue dashed line represents the result of the fitting of the $J_N$ data with Eq. \ref{eq:JN}. Computed values of $J$, $J_N$, $\tau_N$, $\alpha$,  convergence of the $\tau_N$ values and their associated errorbars are reported in the Supplementary Information (SI).} 
\label{fig:taus}
\end{figure*}

\subsection{Survival probability distributions and average transition times in finite size conditions}

As described in section \ref{sec:methods}, sets of 50 to 100 simulations were carried out at five different supersaturation levels $S_1$-$S_5$  (see \ref{tab:setup}). From each set of simulations an \emph{empirical} survival probability distribution (ESP) has been constructed. The average transition time in finite-size conditions $\tau_N$ has been computed for each supersaturation level by a non-linear least square fitting of the ESP with the expression $P_0=\exp\left(-t/\tau_N\right)$, hereafter referred to as the theoretical survival probability (TSP).  
Evaluating the statistical compatibility between the ESP and the TSP with the protocol described in Ref.~\citen{Salvalaglio2014assessing}, allowed ensuring that the crucial hypothesis of negligible bias deposition at the transition state has been satisfactorily fulfilled for all the simulation sets $S_1$-$S_5$. 
In Fig.\ref{fig:taus}a both the ESP constructed from WTmetaD simulations and the fitted TSP are reported; in Fig.\ref{fig:taus}b the supersaturation-dependent values of the finite-size transition time $\tau_N$ is reported as a function of supersaturation, together with the average acceleration factor $\alpha$ (in the inset).  
Fig. \ref{fig:taus}b highlights how the application of WTmetaD allows to directly simulate nucleation events characterised by transition times of the order of $10^4$ seconds, thus significantly expanding the range of transition times that could be reached with standard MD in a similar simulation setup. 

\subsection{Nucleation rates}

%\begin{table*}[ht!]
%\caption{Nucleation average times $\tau_N$, nucleation rates in from NVT simulations $J_N$, rescaled nucleation rates in macroscopic conditions $J$, and average acceleration factor $\alpha$. The errorbars on $\tau_N$ have been computed with a bootstrap-like approach as in Ref. \cite{UnbindingPNAS}}
%\begin{center}
%\begin{tabular}{c|cccc}
%\hline
%Label & ${\tau}_N$ & ${J}_N$& ${J}$&  ${\alpha}$\\
%        & \small{[s]} & \small{[cm$^{-3}$ s$^{-1}$]} &\small{[cm$^{-3}$ s$^{-1}$]} \\
%\hline      
%$S_1$ & 5.75 $\pm$0.65$\times 10^{-8}$ & 1.5$\pm$0.34$\times 10^{25}$& 3.04$\pm$0.70$\times 10^{25}$
%& 2.8 \\
%$S_2$ & 2.33 $\pm$0.33$\times 10^{-5}$  & 2.8$\pm$0.82$\times 10^{22}$& 8.64$\pm$2.53$\times 10^{22}$& 1.8$\times 10^2$ \\
%$S_3$ & 8.02 $\pm$1.96$\times 10^{-2}$  & 6.4$\pm$3.3$\times 10^{18}$& 5.09$\pm$2.65$\times 10^{19}$ & 2.4$\times 10^5$\\
%$S_4$ & 3.61 $\pm$0.76$\times 10^{1}$   & 1.26$\pm$0.56$\times 10^{16}$ & 2.57$\pm$1.14$\times 10^{17}$& 6.3$\times 10^7$\\
%$S_5$ & 3.13 $\pm$0.84$\times 10^{4}$   & 1.30$\pm$0.75$\times 10^{13}$& 1.35$\pm$0.78$\times 10^{15}$ & 1.7$\times 10^{11}$ \\
%\hline
%\end{tabular} 
%\end{center}
%\label{tab:Times}
%\end{table*}
As discussed in section \ref{sec:nucleation_rates} nucleation rates in the confined regime $J_N$ can be directly computed as $J_N=\left(\tau_NV\right)^{-1}$. In Fig. \ref{fig:taus}c values of $J_N$ are reported as a function of the supersaturation $S$, clearly showing a significant extension of the accessible nucleation rates domain, which is increased up to ten orders of magnitude compared to the domain typically accessible by standard molecular dynamics\cite{ArgonRateIsotherms,RegueraRate,WaterTanaka,WaterZipoli,yuhara2015nucleation}.

The surface tension $\gamma$ obtained from the fitting of the $J_N$ values computed from WTmetaD corresponds to $\gamma=16.9$ mN/m, a value that nicely extrapolates the data of Goujon \emph{et al.} \cite{Goujon2014} for the same system in the temperature range between 85 K and 135 K as shown in the Supplementary Information (SI).
The surface tension has been considered independent from $S$, as typically done in CNT. We have found that this choice allows to well describe the $J_N$ data directly computed from simulations while keeping at a minimum the number of fitting parameters. We have also verified that considering $\gamma$ a linear function of supersaturation does not noticeably improve the description of the WTmetaD data. 

\begin{figure}[H]
     \begin{center}
       \includegraphics[width=1\columnwidth]{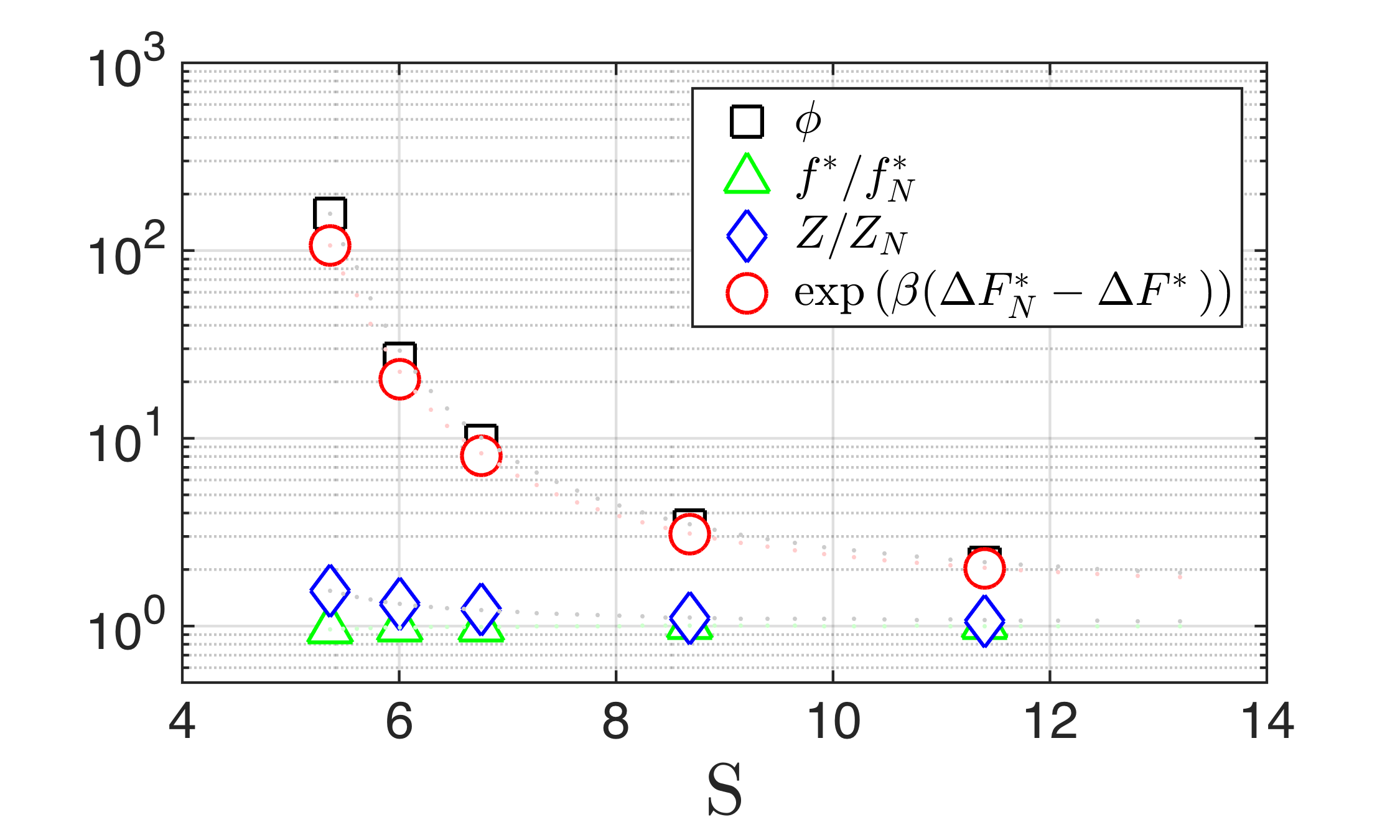}
    \end{center}
   \caption{Breakdown of the contributions to the finite size correction $\phi$ of the factors $\exp\left(\beta(\Delta{F}^*_N-\Delta{F}^*)\right)$, $f/f_N$, and $Z/Z_N$.}  
\label{fig:Zeldovich}
\end{figure}

In Fig. \ref{fig:Zeldovich},  it can be seen that the finite size correction is negligible at high supersaturation ($S\geq{11.4}$), where the nucleation barriers $\Delta{F}^*$ and $\Delta{F}_N^*$ are almost indistinguishable. However, for lower supersaturation levels $\phi$ reaches values accounting for up to two orders of magnitude of difference between $J$ and $J_N$. This finding highlights the importance of properly handling finite size effects when investigating transitions at low supersaturations. A breakdown of the contributions of the terms appearing in $\phi$ is also reported in Fig. \ref{fig:Zeldovich}. It can be seen that, as expected, the contribution of the term $f^*/f_N^*$ negligible over the entire supersaturation domain. Despite the term $Z/Z_N$ having a slightly heavier impact on $\phi$, it can be seen that the finite-size correction is substantially captured by considering only the exponential term in Eq. \ref{eq:ratio}.
Nucleation rates in macroscopic conditions $J$ are thus computed as: 
\begin{equation}
J=\phi{J_N}\simeq\,J_N\exp\left(\beta\left(\Delta{F^*_N}-\Delta{F^*}\right)\right)
\end{equation}
and reported in Fig. \ref{fig:taus}c. Nucleation rates rescaled explicitly accounting also for the term $Z/Z_N$ are reported in the SI. 

\section{Conclusion}
To conclude, in this work we have shown that WTmetaD can be applied to the direct calculation of nucleation rates, proving to be particularly useful to tackle the timescale limitations that plague small-scale nucleation simulations.
In the case of argon condensation this implies being able to simulate nucleation in fairly small systems, while efficiently reaching timescales of the order of $1\times10^4$ s with ordinary computational resources. 
Moreover, we have highlighted that rate calculations from small scale simulations require a systematic assessment of finite-size effects. Being able to simultaneously address both timescale and finite-size limitations allows to significantly extend the range of nucleation conditions that can be directly investigated with computationally efficient, small-scale molecular simulations.

\section*{Acknowledgements}
The authors acknowledge the computational resources provided by the Swiss Center for Scientific Computing (CSCS) and the Brutus Cluster at ETH Zurich. M.P. acknowledges the European Union grant ERC-2009-AdG-247075 and the NCCR MARVEL project for funding.

\bibliography{Condensation_refs}

\end{document}

\title{Overcoming timescale and finite-size limitations to compute nucleation rates from small scale Well Tempered Metadynamics simulations: Supplementary Information.}

\author{Matteo Salvalaglio}
\email{m.salvalaglio@ucl.ac.uk}
\affiliation{Department of Chemical Engineering, University College London, London WC1E 7JE, United Kingdom}
\author{Pratyush Tiwary}
\affiliation{Department of Chemistry, Columbia University, New York, New York 10027, United States of America}
\author{Giovanni Maria Maggioni}
\affiliation{Institute of Process Engineering, ETH Zurich, CH-8092 Zurich, Switzerland}
\author{Marco Mazzotti}
\affiliation{Institute of Process Engineering, ETH Zurich, CH-8092 Zurich, Switzerland}
\author{Michele Parrinello}
\affiliation{Department of Chemistry and Applied Biosciences, ETH Zurich,CH-8092 Zurich, Switzerland}
\affiliation{Facolt\`a di informatica, Istituto di Scienze Computazionali, Universit\`a della Svizzera Italiana, CH-6900 Lugano, Switzerland}

\maketitle

\section{Transition times: Tables and convergence}
Average transition times $\tau_N$, rates in the confined system $J_N$, and corrected, macroscopic rates rates $J$ are reported in Tab. \ref{tab:Times}.
In order to measure the compatibility between the empirical survival probability distribution computed from WTmetaD simulations and the exponential distribution theoretically predicted by the law of rare events a Kolmogorov-Smirnov (KS) test has been carried out \cite{Salvalaglio2014assessing}. 
The p-value associated to the KS statistic is reported in Tab. \ref{tab:Times}. The p-value, in this context, represents the probability that the survival probability distribution obtained from WTmetaD obeys the theoretical exponential distribution. In all cases the p-value is well above the standard significance level of 0.05. In Ref. \citen{Salvalaglio2014assessing} this analysis approach is validated and discussed at length.

\begin{figure}[H]
     \begin{center}
       \includegraphics[width=0.6\columnwidth]{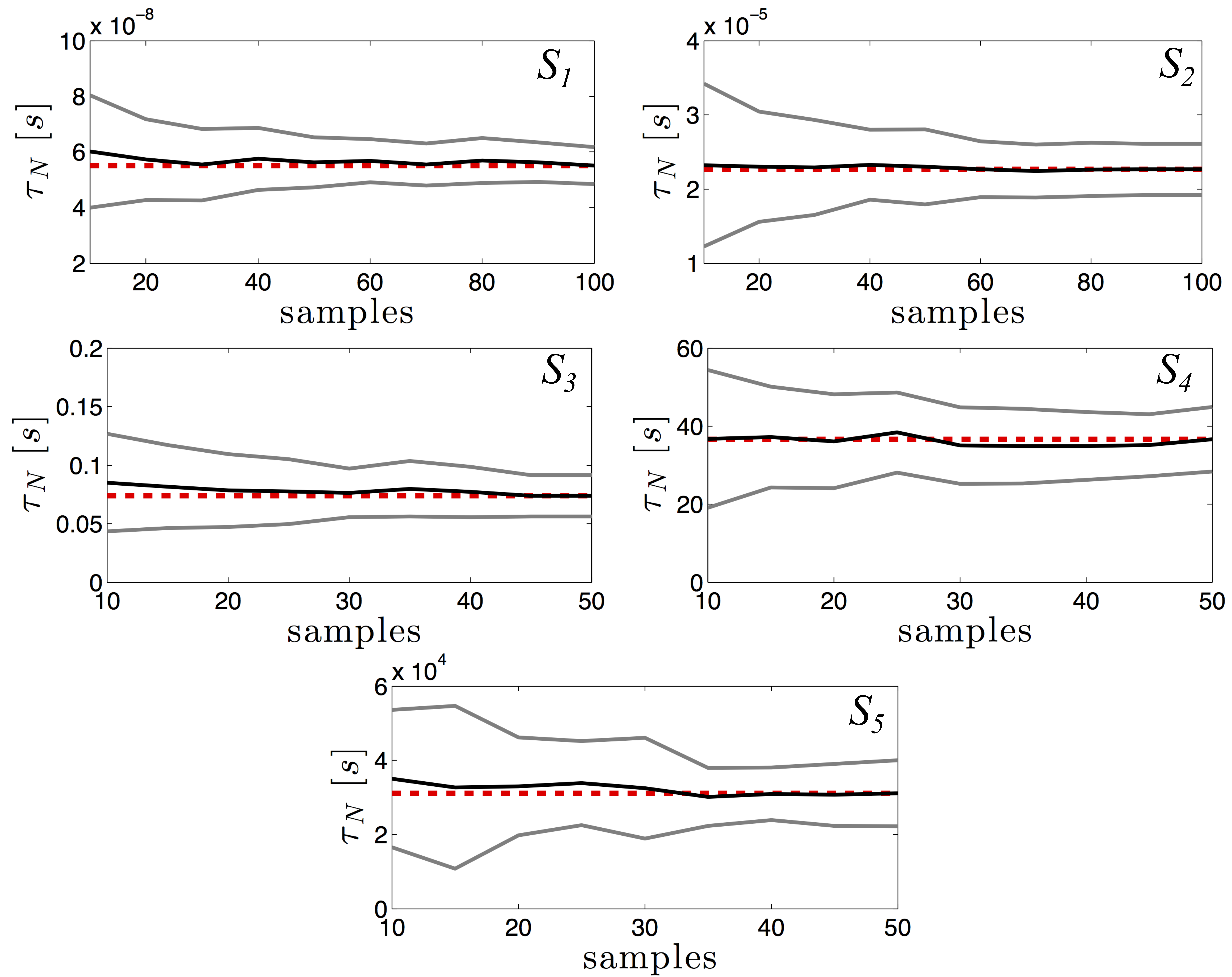}
    \end{center}
   \caption{Convergence of the characteristic time with the number of independent samples.}  
\label{fig:convergence1}
\end{figure}

\begin{table}[H]
\begin{center}
\begin{tabular}{c|cccc|cc}
\hline
Label & $\tau_N$ &  $J_N$& $J$&  $p$-$value$   & ${\alpha}$\\
        & \small{[s]} & \small{[cm$^{-3}$ s$^{-1}$]} &\small{[cm$^{-3}$ s$^{-1}$]} & \\
\hline      
$S_1$ & 5.75 $\pm$0.65$\times 10^{-8}$ &  1.5$\pm$0.34$\times 10^{25}$& 3.04$\pm$0.70$\times 10^{25}$
&  0.84 & 2.8 \\
$S_2$ & 2.33 $\pm$0.33$\times 10^{-5}$ &  2.8$\pm$0.82$\times 10^{22}$& 8.64$\pm$2.53$\times 10^{22}$& 0.67& 1.8$\times 10^2$ \\
$S_3$ & 8.02 $\pm$1.96$\times 10^{-2}$ &  6.4$\pm$3.3$\times 10^{18}$& 5.09$\pm$2.65$\times 10^{19}$& 0.38 & 2.4$\times 10^5$\\
$S_4$ & 3.61 $\pm$0.76$\times 10^{1}$  &  1.26$\pm$0.56$\times 10^{16}$ & 2.57$\pm$1.14$\times 10^{17}$& 0.56& 6.3$\times 10^7$\\
$S_5$ & 3.13 $\pm$0.84$\times 10^{4}$  &  1.30$\pm$0.75$\times 10^{13}$& 1.35$\pm$0.78$\times 10^{15}$& 0.37 & 1.7$\times 10^{11}$ \\
\hline
\end{tabular} 
\end{center}
\caption{Nucleation rates extracted from WTmetaD simulation, $p$-$value$ associated to the Kolmogorov-Smirnov statistic \protect\cite{Salvalaglio2014assessing}, and average acceleration factor.}
\label{tab:Times}
\end{table}
\newpage
The characteristic time convergence with the number of independent samples been reported in Fig. \ref{fig:convergence1}.
Errorbars associated to the characteristic time are computed as the standard deviation of $\overline{\tau}$ obtained from a bootstrap-like analysis.
The convergence of the errorbars as a function of the number of the bootstrap samples is displayed in Fig. \ref{fig:convergence}.
\begin{figure}[H]
     \begin{center}
       \includegraphics[width=0.6\columnwidth]{Convergence.eps}
    \end{center}
   \caption{Convergence of the error estimation as a function of the number of bootstrap samples.}  
\label{fig:convergence}
\end{figure}

\newpage
\section{Importance of the Zeldovich factors ratio}
\begin{figure}[H]
     \begin{center}
\includegraphics[width=0.85\columnwidth]{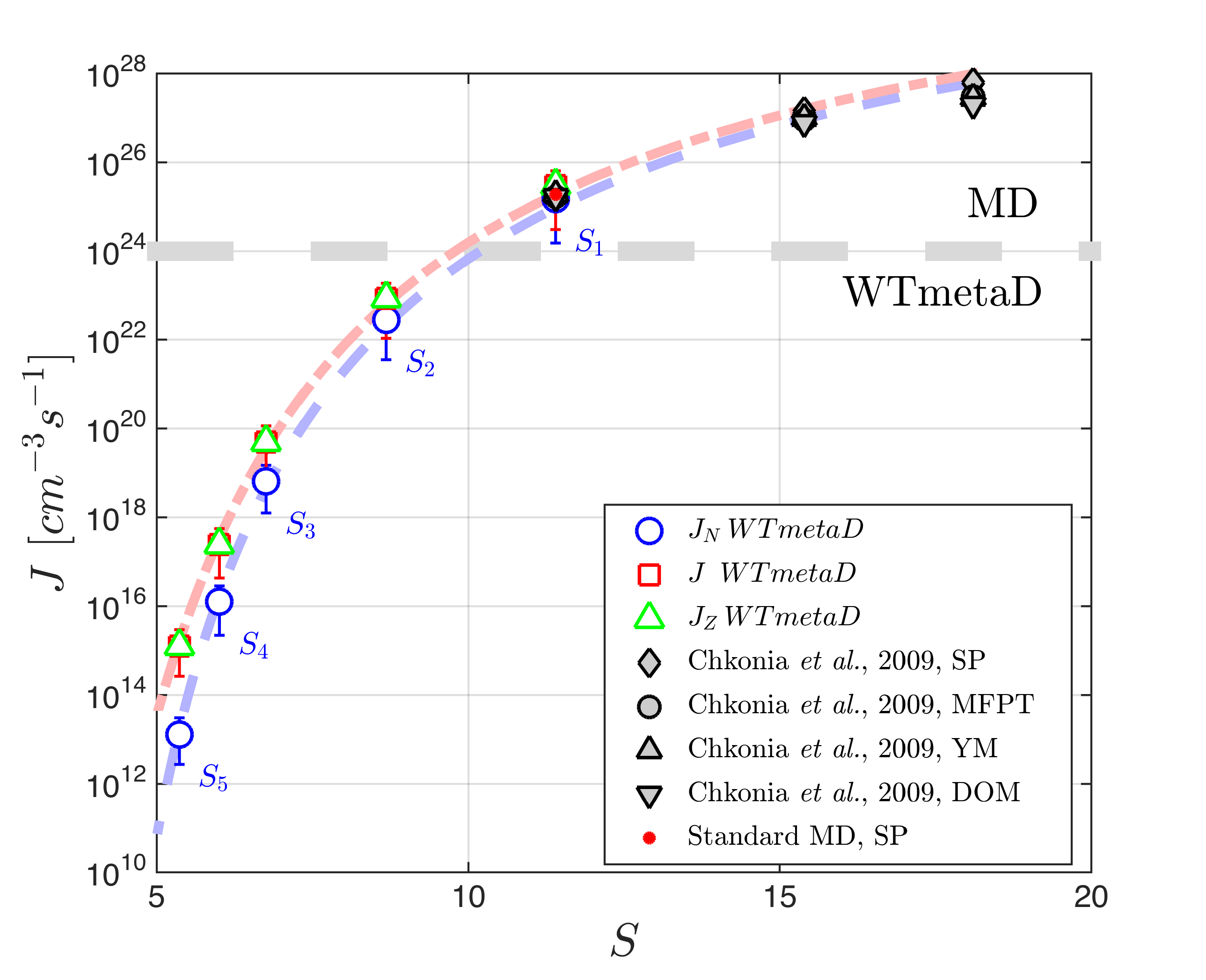}
\caption{Nucleation rates obtained from WTmetaD simulation (${J}_N$), corrected for finite-size effects using Eq. 20 ($J$), and corrected for finite size explicitly accounting for the ratio of Zeldovich factors (${J}_Z$). It can be seen that $J_Z$ values are well within the errorbar of the $J$ values. Literature data\protect\cite{RegueraRate} obtained with different post-processing techniques at high supersaturation has been reported (SP survival probability, MFPT mean first passage time, YM Yasuoka-Matsumoto, DO direct observation) together with an independent calculation of the nucleation rate at S=11.4 performed applying the SP method to standard MD simulations.}  
\end{center} 
\label{fig:Zeldovich_rates}
\end{figure} 

\newpage

\section{Surface tension}

Fitting the WTmetaD characteristic times allows to compute surface tension $\gamma$. In our case the surface tension is $\gamma=16.9$ mN/m. Our estimate agrees well with the trend recently reported in the literature \cite{Goujon2014} for a truncated Lennard-Jones potential. In order to appreciate such agreement, in Fig S4 the $\sigma$ value computed at 80.7 K is reported together with surface tension values of Ref. \citen{Goujon2014}.
\begin{figure}[H]
\begin{center}
\includegraphics[width=0.5\columnwidth]{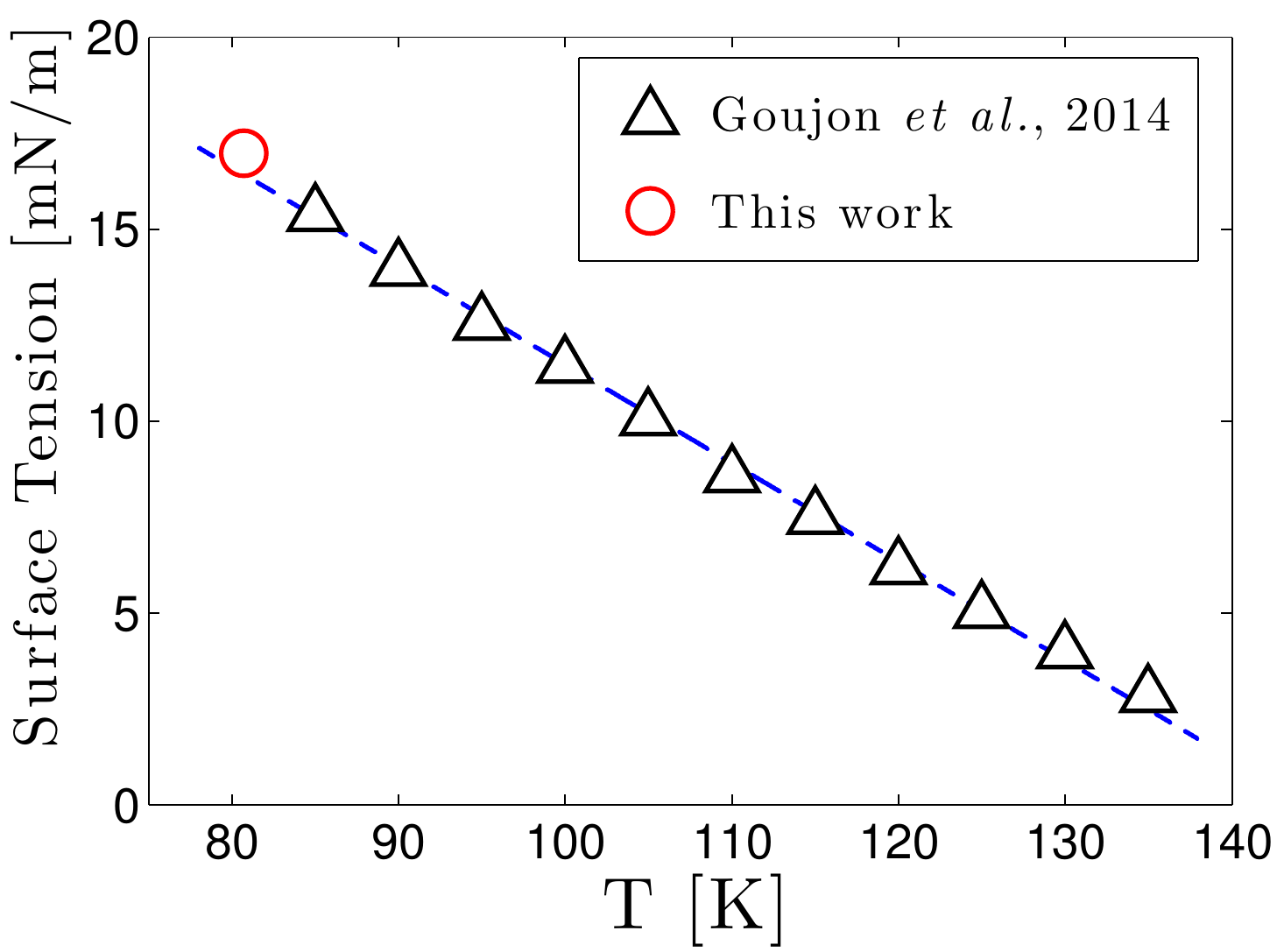}
 \caption{Surface tension as a function of temperature. The value calculated from the fitting of the nucleation rates ${J}$ is reported together with the data from Goujon \emph{et al.} \protect\cite{Goujon2014}. }  
\end{center} 
\end{figure}

\bibliography{Condensation_refs}